\documentstyle[12pt,epsf]{article}
\begin{document}
\centerline{\Large \bf How do we know the charges of quarks?}

\vskip 1cm

\begin{center}

Saurabh D. Rindani

\vskip .5cm
 {\it Theory Group, Physical Research Laboratory
\\

Navrangpura, Ahmedabad 380 009}

\end{center}
\vskip 1cm

 \centerline{\small \bf Abstract}

\vskip .5cm

{\small Experimental tests performed in the past to determine
quark charges are reviewed. It is pointed out that only
experiments involving two real photons can really distinguish
between the Gell-Mann-Zweig scheme of fractional charges from the
Han-Nambu scheme of integer charges in the context of a theory
with gauged colour interactions. While many experiments with two
real photons have been performed, the only experiment with a
definitive statement on quark charges was one on hard scattering
of real photons at CERN. Some theoretical issues not fully
resolved are pointed out. } \vskip 1cm

 One of the questions which interested  Prof. Rajasekaran from
the time he worked on gauge theories in TIFR was the question of
quark charges -- are they fractional (in accordance with the
Gell-Mann-Zweig scheme \cite{GMZ}) or are they integral (in
accordance with the Han-Nambu scheme \cite{HN})? The question was
not easily resolved in the context of a theory with gauged colour
interactions. He continued to work on it when he moved to Madras.

The colour degree of freedom was incorporated in the
Gell-Mann-Zweig (GZ) scheme with a colour triplet of quarks of
each flavour, with all the members having the same (fractional)
charge. In the Han-Nambu (HN) scheme, the members of the colour
triplet were quarks with different charges for the same flavour,
the colour-averaged charge being fractional. Thus, the charge
operator could be written as
\begin{equation}\label{charge}
  Q = Q_0 + Q_8,
\end{equation}
where $Q_0$ is the colour-singlet charge, viz., $\frac{2}{3}$ for
up quarks, and $-\frac{1}{3}$ for down quarks, and $Q_8$ is the
colour contribution to the total charge $Q$, which vanishes in the
GZ scheme, and is given in the HN scheme by
\begin{equation}\label{octet}
  Q_8 = \left( \begin{array}{ccc} \frac{1}{3} & 0& 0\\
                                    0 & \frac{1}{3} & 0 \\
                                    0 & 0 & -\frac{2}{3}
                                    \end{array} \right),
\end{equation}
where the matrix $Q_8$ is in colour space. The presence of $Q_8$
gives rise to the charges (1,1,0) for the members of the colour
triplet for each up quark, and (0,0,$-1$) for each down quark.

At first sight it appears that it would be easy to decide between
the two models in experiments like deep-inelastic scattering which
use electromagnetic probes. However the first requirement in these
cases is that the relevant energy should be above the 
threshold for excitation of colour \cite{lipkin1}. Below the
colour threshold, only $Q_0$ would be effective, giving no
difference between the predictions of the two models.

It was thus believed that deep-inelastic scattering experiments,
which presumably had high enough energies to excite colour, would
reveal the true quark charges. However, it was shown by
Rajasekaran and Roy \cite{RR}, and Pati and Salam \cite{pati},
that in a theory in which colour is gauged \cite{PS}, colour breaking
induces mixing between the electroweak and colour gauge fields. As
a result, the effective charge operator for large squared
momentum-transfer takes the form
\begin{equation}\label{qeff}
Q_{\rm eff} (q^2) = Q_0 + Q_8 \left(
\frac{-m_g^2}{q^2-m_g^2}\right).
\end{equation}
Thus for $\vert q^2 \vert >> m_g^2$, $Q_{\rm eff} \approx Q_0$. It
follows that no difference between the two models would be seen
asymptotically in deep-inelastic scattering (DIS), so far as quark
charges are concerned. However, gluons in the HN scheme carry
electric charge. Hence they would also contribute to the DIS
structure functions, and if this effect is large, detecting it
would signal integrally charged quarks.

It is curious that even though the longitudinal polarization
vector of massive spin-one gluons goes for large $\vert q \vert$
as $q^{\mu}/m_g$, the gluon contribution to the structure function
$F_2$ is finite for $Q^2 \rightarrow \infty $ \cite{RR}; it is
however suppressed.

With the advent of high energy $e^+e^-$ colliding beams at PETRA
there was a possibility of looking for direct production of
charged-gluon jets. With the assumption that colour was excited at
these energies, the additional gluon \cite{2jet} and coloured
scalar contributions \cite{scalars} was expected to provide a test
of the HN model in two-jet production experiments. In particular,
the angular distribution of jets is different in the HN model as
compared to the $1+\cos^2\theta$ distribution in the GZ model
\cite{2jet, scalars}. This, to my knowledge, was never tested by
experimentalists. The three-jet experiments, in spite of a scaling
violation contribution of three-gluon production in the HN model,
merely provided a limit on the gluon mass \cite{3jets}.

The problem of nonobservation of the octet charge for large $\vert
q^2 \vert$ may be avoided in two-photon experiments, first carried
out at PETRA. For, in such experiments, the relevant observable is
$Q_{\rm eff}^2$, which has a contribution $Q_8^2$. Hence even
below colour threshold, a colour singlet projection of $Q_8^2$
would survive, and would help discriminating between the two
charge schemes. The problem, however, was that in PETRA
experiments the $\vert q^2 \vert$ values were never close to zero,
and the experiments could accommodate the GZ as well as HN models,
for a range of values of $m_g$ \cite{Petra}.

In the case of single photon production in $e^+e^- \rightarrow 2$
jets, there would be contributions from photons radiated off
final-state quarks (and also gluons in the case of HN scheme), as
well as from initial state $e^+$ or $e^-$. The contribution coming
from final-state radiation would  measure the sum of the squares
of quark or gluon effective charges. However, since one of the
photons involved is virtual, the resultant effect is not very
sensitive to the model \cite{eephoton}. Some experimental papers
do attempt to make a comparison, without conclusive results
\cite{eephotonexp}.

It seemed clear that the best test of quark charges would be in
processes where two strictly real photons were involved. However,
high energy processes involving real photons are plagued with
uncertainties due to large errors.

To date, an inexhaustive list of various experiments dealing with
two real photons is:
\begin{enumerate}
  \item  LEP (ALEPH, DELPHI, L3 and OPAL collaborations): $e^+e^-
  \rightarrow \gamma\gamma + X$ \cite{LEP}.
  \item NA3 spectrometer: $\pi^{\pm} + C  \rightarrow \gamma\gamma
  +X$ (200 GeV/$c$) \cite{NA3}.
  \item AFS: $pp \rightarrow\gamma\gamma
  +X$ \cite{AFS}.
  \item Tevatron (CDF and D0 collaborations): $p\bar p \rightarrow \gamma\gamma
  +X$ \cite{Tev}.
  \item E-706 (Fermilab): $\pi^{\pm},p + N ({\rm Be,Cu,H})
  \rightarrow \gamma\gamma
  +X$ (515 GeV/$c$ $\pi^{\pm}$ or 530 and 800 GeV/$c$ $p$) \cite{E706}.
  \item NA14: $\gamma N ({\rm Li}^6)\rightarrow \gamma X$ \cite{NA14,NA142}.
\end{enumerate}

A good experimental test would be in photon-pair production in
$e^+e^- \rightarrow \gamma\gamma + 2$ jets. In this case, the
colour-singlet projection of the square of octet charges would
contribute in the case of HN, and a quantity
\begin{equation}\label{q3ext}
  \sum_{\rm flavour} \vert \langle Q_0Q_8Q_8 \rangle_{\rm colour~
  ave.} \vert ^2
\end{equation}
would be measurable. The effective charge factor
\begin{equation}\label{q3eff}
  \sum_{\rm flavour} \vert \langle Q^3 \rangle_{\rm colour~
  ave.} \vert ^2
\end{equation}
amounts to $11/9$ in HN as compared to $131/243$ in GZ, for five
flavours). Data does exist; however no comparison has been
attempted.

Comparison of experimental results with theoretical predictions in
HN and GZ schemes have been made in some hadronic experiments
\cite{pp2g}. However, one has to have some reservations about
these because of the following factors. (i) Uncertainties in quark
and gluon distributions. Particularly, since gluons could also
contribute to DIS, even in the zeroth
    order in $\alpha_{\rm QCD}$, their distributions as determined from a fit
to DIS experimental data would be different in HN and GZ schemes.
(ii) Higher-order corrections in HN model are not known, and a
simple ``$K$ factor" calculated for GZ scheme has to be assumed.

The experiment which has led to a definitive statement, modulo the
assumptions already mentioned, is the NA14 experiment on hard
scattering of real photons \cite{NA14,NA142}. Here the prediction in HN
model is larger than that in the GZ model by a factor of 2.65 for
four flavours (10/3 in HN scheme as against 34/27 in GZ scheme).
The HN Born contribution has been simply scaled up by the same QCD
$K$ factor as in the GZ contribution. HN model seems disfavoured
by at least two standard deviations. Fig. 1, which is taken from
\cite{NA14}, shows a comparison of the data with theoretical
expectations from the two schemes. Note that the gluon
contribution is not included, or else the factor of 2.65 could be
replaced by something larger.
\begin{figure}[hbt]
\leavevmode
  \begin{center}
  \epsfysize 14cm
  \epsfbox{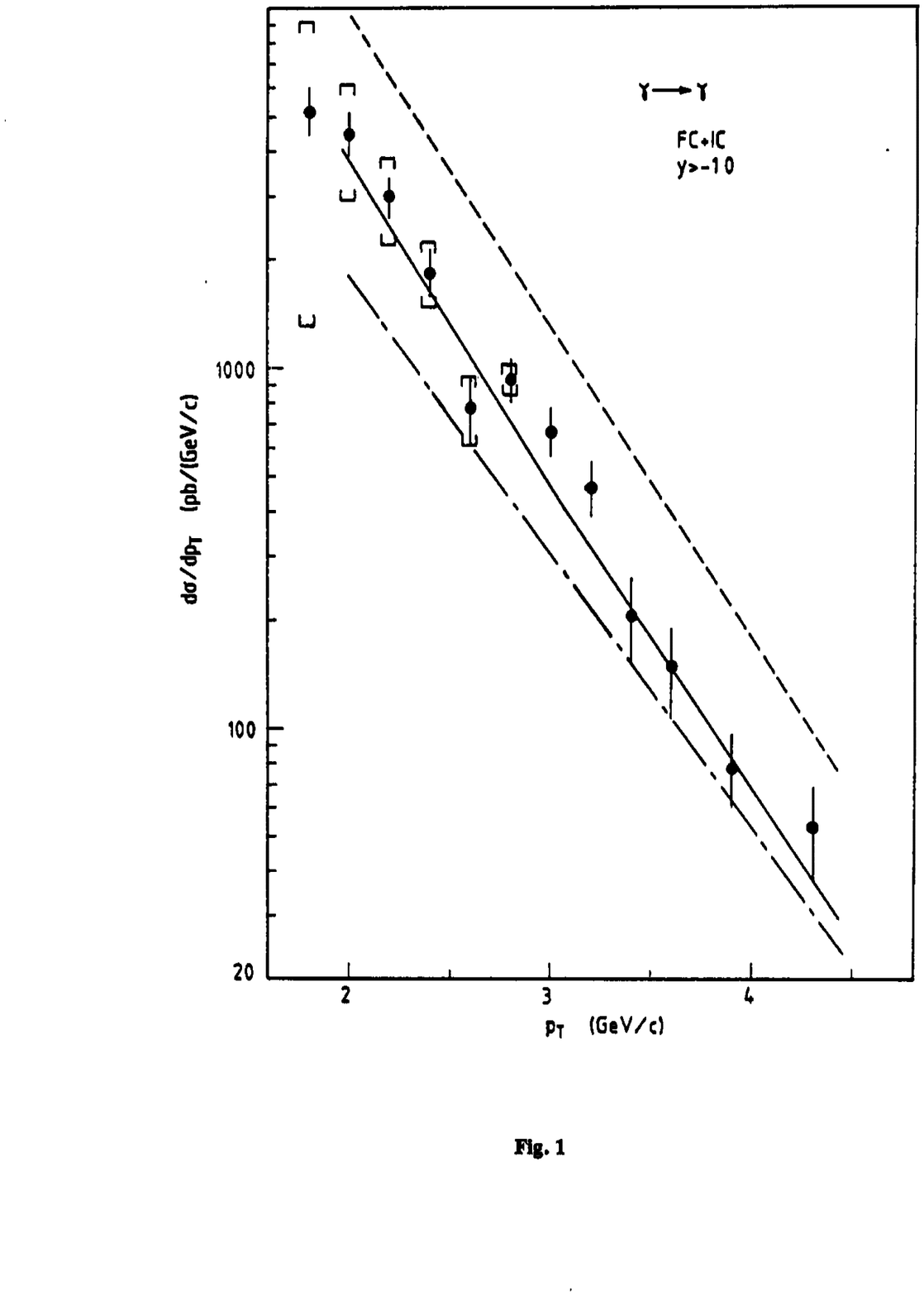}
  \end{center}
  \caption{\small The $p_T$ distribution of inclusive prompt photons from the 
NA14 experiment \cite{NA14}. The dash-dotted and solid curves correspond
respectively to the
Born term and the Born term  together with next-to-leading order QCD 
corrections in GZ
model. The dashed curve is 2.65 times the solid curve, 
the expectation from the gauged HN model.
  }\label{}
\end{figure}

 Present day results from Tevatron have the potential for
making more definitive statements. However, no comparison has been
made.

Other experiments where some comparison could be made is radiative
decays of hadrons, particularly, two-photon decays of mesons
\cite{field}.

Theoretical topics which have not been sufficiently investigated
are (i) possible phenomenon of colour oscillations \cite{lipkin2}
(ii) effects that might be seen at colour threshold (iii) masses
and spectrum of coloured Higgs scalars, and the possibility of
colour singlet bound states of these among themselves or with
quarks (for a discussion on a similar problem with unconfined
fractionally charged quarks, see \cite{okun}).

In conclusion, data from NA14 experiment on hard scattering of
real photons seems to indicate that the HN model is strongly
disfavoured. However, this depends on some assumptions stated
earlier. $e^+e^-$ experimental data from LEP could be more
definitive. Some theoretical issues like those of colour
oscillations, however, need to be addresses in that context.

I thank Prof. Rajasekaran for a critical reading of the manuscript.

\end{document}